\documentclass[aps,prb,twocolumn,superscriptaddress,groupedaddress]{revtex4}
\usepackage{epsfig,amsopn}
\usepackage{graphicx}
\usepackage{amsmath,amssymb}
\usepackage{physics}
\usepackage{enumerate}
\usepackage{sidecap}
\usepackage{wrapfig}
\usepackage[usenames]{color}
\newcommand\bea{\begin{eqnarray}}
\newcommand\eea{\end{eqnarray}}
\newcommand\beq{\begin{equation}}
\newcommand\eeq{\end{equation}}

\newcommand{\bib}{\bibitem}
\def\nn{\nonumber}
\def\f{\frac}

\def\ep{\epsilon}

\def\lam{\lambda}
\def\si{\sigma}

\def\De{\Delta}
\def\dg{\dagger}

\def\ka{\kappa}
\def\th{\theta}

\def\tb{t_{\bot}}

\begin{document}
\title{Enhanced specular Andreev reflection in bilayer graphene}
\author{Abhiram Soori$^{1,2}$, Manas Ranjan Sahu$^{1}$, Anindya Das$^{1}$, and Subroto Mukerjee$^{1}$\\
{\it $^1$ Department of Physics, Indian Institute of Science, Bengaluru 560012, India. \\
$^2$ International Centre for Theoretical Sciences, Survey No. 151,
 Tata Institute of Fundamental Research,
 Shivakote,  Hesaraghatta Hobli, Bengaluru 560089, India.}
}
\begin{abstract} 
Andreev reflection in graphene is special since it can be of two 
types, retro or specular. Specular Andreev reflection~(SAR) dominates when 
the position of the Fermi energy in graphene is comparable to
or smaller than the superconducting gap. Bilayer graphene~(BLG) is an 
ideal candidate to observe the crossover from retro to specular since 
the Fermi energy broadening near the Dirac point is much weaker compared
to monolayer graphene. Recently, the observation of signatures of SAR in BLG have been 
reported experimentally by looking at the enhancement of conductance at 
finite bias near the Dirac point. However, the signatures were not very pronounced 
possibly due to the participation of normal quasiparticles at bias energies close to the superconducting gap. 
Here, we propose a scheme to observe the features of enhanced SAR even at zero bias at a normal
metal~(NM)-superconductor~(SC) junction on BLG. Our scheme involves applying
a Zeeman field to the NM side of the NM-SC junction on BLG (making the NM ferromagnetic), which 
energetically separates the Dirac points for up-spin and down-spin. 
We calculate the conductance as a function of chemical potential and bias 
within the superconducting gap and show that well-defined regions of 
specular- and retro-type Andreev reflection exist. We compare the results with and without 
superconductivity.  We also investigate the 
possibility of the formation of a p-n junction at the interface between the
NM and SC due to a work function mismatch. 
\end{abstract}
\maketitle
\section{Introduction}
Andreev reflection~(AR) - a scattering process by which a current 
can be driven into a superconductor~(SC) from a normal metal~(NM) 
by applying a bias within the superconducting gap - was first 
discovered by Andreev~\cite{andr64} and has been extensively 
studied for several decades~\cite{btk,kasta}. Graphene on the 
other hand has attracted a huge interest in the past decade owing to 
its electronic and material properties~\cite{novoselov04,neto09,rozh16,mccan13}.
Graphene is a semimetal whose electronic structure can be described by
a Dirac Hamiltonian (with a vanishingly small mass). Andreev reflection has been
studied both theoretically~\cite{bhatta06,beenak06,benjamin08,rainis09,majidi12} 
and experimentally~\cite{sahu16} in graphene. What makes Andreev 
reflection in graphene special is that it can be of two types: 
one where the reflected hole retraces the path of the incident electron 
(called retro-) and another where the reflected hole moves away not tracing 
back the path of the incident electron (called specular-)~\cite{beenak06,sahu16}.
Specular Andreev reflection has not been observed in graphene due to 
charge density fluctuations across the sample~\cite{sahu16}, but a weak qualitative 
agreement is observed in bilayer graphene~\cite{efet16,efet16-prb}. Bilayer
graphene~(BLG)~\cite{ludwig07} is a better 
candidate to observe specular Andreev reflection since charge density
fluctuations are much smaller than in monolayer graphene. In the experimental setup,
a part of the BLG is kept in proximity to a SC, which 
induces superconducting correlations on BLG. It can be seen in 
Fig.~3(a) of  Ref.~\cite{efet16} which shows only a  
weak qualitative agreement between the experimental observations and 
underlying theoretical calculations (note also the very different color scales 
of the experimental and theoretical plots required to arrive at even this level
of agreement).

 Generally speaking, Andreev reflection is a process
where an electron incident from a normal metal into the superconductor results in 
a reflected hole. This is equivalent to saying that two electrons on the normal
metal side- one from above the Fermi energy and one from below the Fermi energy 
pair up and go into the superconductor as a Cooper pair~\cite{btk}. We
use the latter convention for our analysis.

\begin{figure}[htb]
\includegraphics[width=8cm]{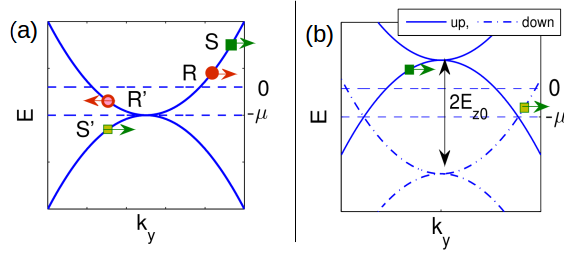} 
 \caption{The subgap bandstructures of the NM part of the NM-SC setup. 
 (a) Zero Zeeman field in the NM part. The points
 $R$ and $R'$ correspond to two electrons contributing to retro Andreev reflection,
 while the points $S$ and $S'$ correspond to electrons contributing to specular
 Andreev reflection. (b) Finite Zeeman field $E_{z0}$ in the NM part. The dispersion
 for up-spin and down-spin have the CNP's separated well energetically. 
 Both the electron states  shown contribute to specular Andreev reflection.  
 }~\label{fig-schem}
\end{figure}

In a manner similar to that for Andreev reflection in monolayer graphene~\cite{beenak06}, 
retro- and specular- Andreev reflection can also be understood in bilayer 
graphene~\cite{efet16,efet16-prb}. If both the electrons participating in the reflection 
come from the same side of the charge neutrality point~(CNP), the Andreev reflection is of 
the retro type, while if the two electrons come from opposite sides of the CNP,
the Andreev reflection is of the specular type. This is because, the momentum
of the reflected hole along the $y$-direction has to be same as that of the incident electron. This 
means that when the hole originates from the same side of the CNP as that of the incident electron, 
the velocities along the $y$-direction of the two electrons participating in Andreev reflection
have opposite signs. On the other hand, when the hole originates from the opposite side of the CNP 
as that of the incident electron, the velocities along the $y$-direction of the two electrons 
participating in Andreev reflection have the same sign. This is shown in
Fig.~\ref{fig-schem}(a). 

Furthermore, the two electrons must have opposite spin. This allows us to separate the
CNPs for the up-spin and the down-spin bands by applying a 
Zeeman field. In this work, we add a Zeeman field $E_{z0}$ to the NM part of the NM-SC junction
on BLG and calculate the conductance spectrum as a function of chemical potential and bias energy.
As shown in Fig.~\ref{fig-schem}(b), for small chemical potential ($|\mu|<E_{z0}$) and small
bias ($|eV_{bias}|<E_{z0}-|\mu|$) the Andreev reflection is specular. 
We discuss several features of the conductance spectrum in the presence of a Zeeman field, 
where the main highlight is the enhanced specular Andreev reflection~(SAR) at zero chemical
potential and zero bias energy. 

The paper is organized as follows. In Sec.~\ref{sec-calc}, the calculation is presented.
In Sec.~\ref{sec-result}, we  show the main results. In Sec.~\ref{sec-scnm}, a comparative 
analysis replacing the superconductor with normal metal is discussed. In Sec.~\ref{sec-expt},
connection to experiments is discussed. Finally, in Sec.~\ref{sec-summary}, the work is summarized.
In Appendix~A, calculations for the system where the superconductor is replaced with normal metal 
are shown. In Appendix~B, the system where the effect of step height is extended in the normal 
metal region is studied. 
 
\section{Calculation}~\label{sec-calc}
The BLG Hamiltonian at either of the two degeneracy points is: 
\beq H_0 = \hbar v (k_x \si_x-k_y\si_y\lam_z)
-\tb (\lam_x+\lam_x\si_z)/2, \eeq
where $\vec k=(k_x,k_y)$ is the momentum with respect to the $\vec K$
point at the top layer and for the bottom layer, $\vec k=(k_x,k_y)$
is the momentum with respect to $\vec K'$,
$v$ is the Fermi velocity and $\tb$ is the 
coupling between the two layers. The layer
asymmetry term is absent in this Hamiltonian.  
The choice of basis is  $[u_{A1},u_{A2},u_{B1},u_{B2}]$. $A$ and $B$ 
refer to two kinds of lattice points in each layer of graphene, while $1$ and $2$
refer to the two layers of graphene. $\si$'s are the Pauli matrices in the
$A,B$-basis, while $\lambda$'s are the Pauli matrices in the $1,2$-basis. 
This Hamiltonian can be  diagonalized to get the eigenspectrum 
$E(\vec k)=\nu_{\si}\sqrt{(\hbar v \vec k)^2+\tb^2/2 + 
\nu_{\lam}\tb \sqrt{(\hbar v \vec k)^2+\tb^2/4 }}$, 
where $\nu_{\lam},~\nu_{\si}=\pm1$. The index $\si$ corresponds to the
bipartite pseudospin in graphene and the index $\lam$  corresponds to the
two layers of BLG. 

The eigenvector at an energy $E$ and momentum $(k_x,k_y)$ is: 
\bea
\vec u(E,k_x) &=& \f{1}{N}\begin{bmatrix}
            -\tb E^2 \\
            [E^2-(\hbar v \vec k)^2]E \\
            -\tb \hbar v k_-E \\
            \hbar v k_+ [E^2-(\hbar v \vec k)^2]
           \end{bmatrix}, ~\label{eigsp}
\eea
where $k_{\pm}=k_x\pm i k_y$ and $N$ is the normalization 
factor for the pseudospin such that $\vec u^{\dg} \vec u =1$.

The Hamiltonian for the NM-SC junction on BLG is:
\beq 
H = [H_0-\mu-U(x)]\tau_z - E_z(x) s_z  +\De(x) \tau_x, \label{ham}
\eeq
where $U(x)= U_0 \eta(-x)$, $s_z$ corresponds to the real spin, 
$E_z(x)= E_{z0}\eta(x)$ is the Zeeman field and can be nonzero 
only on the NM side, $\De(x)=\De\eta(-x)$, $\eta(x)$ is the Heaviside 
step function, and the $\tau$-matrices act in the particle-hole
sector. The wavefunction for an electron at energy $E$
(in the range: $|E|<\De\ll\tb$)
and spin~$s$ 
($s=\pm1$ is the eigenvalue of the operator $s_z$), incident from 
the NM side onto the SC has the form $\psi_s(x)e^{ik_y y}$, such that  

\bea  
\psi_s(x) &=& \Big(e^{-ik^e_xx}~ \vec u_{N,s}(\ep,-k^e_x) + r_N~ e^{ik^e_xx}
~\vec u_{N,s}(\ep,k^e_x)\Big)  \begin{bmatrix}
                          1 \\0
                         \end{bmatrix} \nn \\
& & + ~r_A ~e^{-i k^h_x x}~\vec v_{N,s}(\ep_h,-k^h_x) \begin{bmatrix}
                                             0 \\ 1
                                            \end{bmatrix} \nn \\
& &    + ~\tilde r_N ~e^{-\kappa x}~\vec u_{N,s}(\ep,i\kappa) \begin{bmatrix}
                                             1 \\ 0
                                            \end{bmatrix} \nn \\
& & + ~\tilde r_A ~e^{-\ka^h x}~\vec v_{N,s}(\ep_h,i\ka^h) \begin{bmatrix}
                                             0 \\ 1
                                            \end{bmatrix},~~
                                            {\rm for}~~x>0,
\nn \\ &=& \sum_{j=1}^4 w_{j,s}~  e^{i k^S_j x}~ \vec u_{S}
(k^{S}_j),~~ {\rm for}~~x<0, \eea
where $\vec u_{N,s}(\tilde\ep,k_x)$ and $\vec v_{N,s}(\tilde\ep,k_x)$ are 
the electron and hole sector eigenspinors of the Hamiltonian on
the NM side [given by Eq.~\eqref{eigsp}] with $x$-component of momentum $k_x$,
and $\vec u_S(k^S_j)$ is the eigenspinor
on the SC side with $x$-component of  momentum~$k^S_j$; furthermore,  
the $x$-component of the electron and hole momenta on the NM side are given by:
\bea
\hbar v k^e_x &=& sign(\ep) \sqrt{\ep^2+2\tb|\ep|-(\hbar v k_y)^2} \nn \\
\hbar v k^h_x &=& sign(\ep_h) \sqrt{\ep^2_h+2\tb|\ep_h| -(\hbar v k_y)^2} \nn \\
\hbar v \ka &=& \sqrt{(\hbar v k_y)^2 + 2\tb|\ep|-\ep^2} \nn \\
\hbar v \ka^h &=& \sqrt{(\hbar v k_y)^2 + 2\tb|\ep_h|-\ep_h^2} ,
\eea
where $\ep = (E+\mu+sE_{z0})$ and $\ep_h=(\mu-sE_{z0}-E)$. On the SC side,
$k^S_j$ has a nonzero imaginary part at subgap energies. The complex values
of $k^S_j$  arise as complex conjugates and thus there are 
eight in all. Normalizability allows only four modes (out of
eight) which have a negative imaginary part. Different values of $k^S_j$ are
obtained numerically from the eigenvalue-eigenvector equation. 
We shall employ the boundary condition that the wavefunction is
continuous at $x=0$ to solve for the scattering coefficients. 

{\it Current operator and the conductance}~:
From the Hamiltonian, it can be shown that the
current for the NM part of the BLG has the form
$\vec J_s = e v \psi_s^{\dg} (\si_x,-\si_y\lam_z)\psi_s$. 
The differential conductance  is obtained by summing over $J_s$
for all possible values of $(k_x,k_y)$ and $s=\pm 1$ at a given energy $E$
such that the $x$-component of the velocity of the incident 
electron points along the $-\hat x$ direction. We calculate the 
scattering amplitudes and the conductance of the junction. 
The cross terms ($r_Nr_A$) drop out while calculating the 
conductance and only the terms proportional to $|r_N|^2$ and $|r_A|^2$
contribute to the current. The total 
current is $\vec I = \int dk_x \int dk_y \sum_s \vec J_s(k_x,k_y)$. 
and the only nonzero component of $\vec I$ is along $-\hat x$
(i.e., $\vec I = -\hat x \cdot I$). We are interested in 
calculating the conductance $G=dI/dV$,  which is given by the expression~\cite{landau-butti}
\bea G 
&=& \f{2e^2}{h} \sum_s \f{W (\mu+sE_{z0}+E+\tb/2)}{h v}
\int_{-\th_{c,s}}^{\th_{c,s}} d\th 
~\psi_s^{\dg} \si_x  \psi_s, \nn \\ \label{eq-conductance} \eea  
where $W$ is the width of the bilayer graphene-superconductor interface
and the factor of $2$ is for valley degeneracy.
The critical angle for spin $s$ is given by $\th_{c,s}
=\sin^{-1}{[\min{\{(k_{h,\bar{s}}/k_{e,s}),1}\}]}$ 
where $k_{h,\bar{s}}$ and $k_{e,s}$ are the magnitudes 
of the momenta $\vec k$ in the hole band with spin $\bar{s}$ and
the electron band with spin $s$  ($\bar{s}$ is opposite to $s$)
at energy $E=eV_{bias}$ respectively.

\section{Results}~\label{sec-result}
Results of the conductance calculation for two choices of parameters 
have been plotted as contour plots in Figs.~\ref{fig-results-2}~(a)
and~\ref{fig-results-2}~(b). We discuss the 
features observed in the contour plots below.

{\it Zero Zeeman field}~:
In Fig.~\ref{fig-results-2}~(a), a dominant feature is two dark-thick lines 
that appear along the  diagonals: $eV_{bias}=\pm \mu$. These correspond to 
one of the two electron Fermi surfaces participating in Andreev reflection
at $eV_{bias}=\pm \mu$ having zero circumference. The lines 
$eV_{bias}=\pm \mu$ correspond to crossover from retro- to specular- Andreev
reflection. Another feature is that there are two islands of light-blue color
around $\mu=0,eV_{bias}\sim \pm 0.8 \De$. This corresponds to specular Andreev
reflection since the two electrons participating in the Andreev reflection come 
from above and below  the CNP. All the data-points in
the region $|eV_{bias}|>|\mu|$ correspond to specular Andreev reflection. 
Similarly, all the data points in the region $|eV_{bias}|<|\mu|$ correspond
to retro Andreev reflection. We also notice an asymmetry in $\mu\to -\mu$, 
which is due to a finite $U_0$. These results and the discussion
agree with that in Ref.~\cite{efet16-prb}.

{\it Nonzero Zeeman field}~:
\begin{figure}[htp]
\includegraphics[width=9cm]{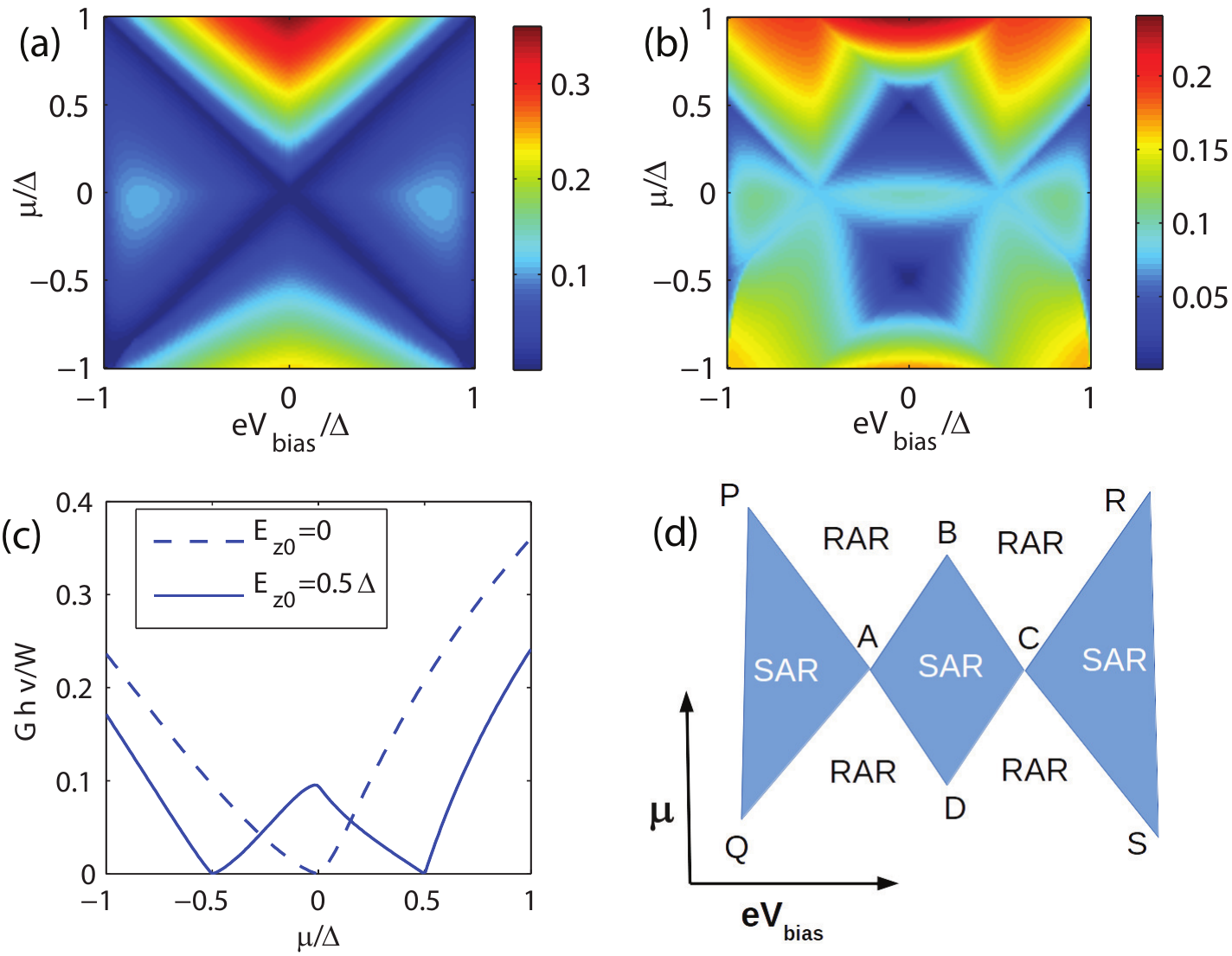}
 \caption{(a-c): $Gh v/ W$ in units of $\tb 2 e^2/h$ is plotted.
  Parameters.- (a):~$\De=0.003\tb$,~$U_0=\De$ and  $E_{z0}=0$. 
  (b):~$\De=0.003\tb$,~$U_0=\De$ and  $E_{z0}=0.5\Delta$. 
 (c): $\De=0.003\tb$, $U_0=\De$ and $eV_{bias}=0$.
(d): Schematic diagram showing regions of specular~(SAR) and retro~(RAR)
Andreev reflections. 
 }~\label{fig-results-2}
\end{figure}
In Fig.~\ref{fig-results-2}~(b), the Zeeman field in the normal metal region
$E_{z0}$ is chosen to be $0.5\De$. The striking features of this contour plot
are: (i) three light blue islands, two of which are located around 
$\mu=0,eV_{bias}\sim \pm 0.8 \De$ and one located around $\mu=0,eV_{bias}\sim 0$,
and (ii) two dark blue patches located around $\mu=0.5\De, eV_{bias}\sim 0$.

To understand the features of Fig.~\ref{fig-results-2}~(b), let us define
different points on the contour plot: $A=(-0.5\De,0)$, $B=(0,0.5\De)$,
$C=(0.5\De,0)$, $D=(0,-0.5\De)$, $P=(-\De,0.5\De)$, $Q=(-\De,-0.5\De)$, 
$R=(\De,0.5\De)$ and $S=(\De,-0.5\De)$ [each of these points is written 
in the form $(eV_{bias},\mu)$ ]. Now, within the diamond $ABCDA$, both the electrons
contributing to Andreev reflection lie on different sides of the charge neutrality
point. So,  Andreev reflection  is specular within this diamond. 
Also, in the triangles $PAQ$ and $RCS$ the two electrons contributing to Andreev
reflection lie on different sides of the CNP. Hence,  Andreev
reflection is specular in these regions. Outside of the two triangles and the diamond, the
two electrons contributing to Andreev reflection lie on the same side of the charge neutrality
point. Hence, in these regions,  Andreev reflection is retro. In each of the two 
 dark blue patches around the points $B$ and $D$ the data points are in proximity
 to CNP for both the electrons participating in the Andreev reflection. 
 Since the size of the Fermi surface approaches zero as one tends to the CNP,
 the conductance is suppressed around points $B$ and $D$. In contrast, along the lines $PA$,
$QA$, $AB$, $BC$, $CD$, $DA$, $RC$ and $CS$ away from the points $B$ and $D$, data points 
for only one of the two participating electrons (in Andreev reflection) is {at the} charge neutrality
point. 

More generally, for a given choice of $E_{z0}$, the diamond $ABCDA$ is formed by the points
$A=(-E_{z0},0)$, $B=(0,E_{z0})$, $C=(E_{z0},0)$, and $D=(0,-E_{z0})$, and the points 
$P=(-\De,\De-E_{z0})$, $Q=(-\De,-\De+E_{z0})$, $R=(\De,\De-E_{z0})$, and 
$S=(\De,-\De+E_{z0})$ form the triangles $PAQ$ and $RCS$. Hence, in the case when
$E_{z0}=0$, the diamond $ABCDA$ has zero area as can be seen in Fig.~\ref{fig-results-2}~(a).
And the regions inside the two triangles $PAQ$ and $RCS$ are described by the inequalities
$-(eV_{bias}+E_{z0})>|\mu|$ and $(eV_{bias}-E_{z0})>|\mu|$, respectively. 
These are the regions where the Andreev reflection is specular. Outside  these regions, the
Andreev reflection is  retro.

Zero bias cuts of Figs.~\ref{fig-results-2}(a) and~\ref{fig-results-2}(b) have been plotted
in Fig.~\ref{fig-results-2}~(c). These clearly show that around the CNP, the zero-bias
conductance is enhanced under an applied Zeeman field, while in the case of zero Zeeman field, 
the zero bias conductance is suppressed. 

{\it Choice of the parameter $U_0$}~:
Previously, we chose $U_0=\De$ so as to allow for significant conductance
despite accounting for a work function mismatch [modeled by the step function $U(x)$]. Now,
we examine the features of the conductance spectrum for different choices
of $U_0$ and make a connection to previous works. 

\begin{figure}[h]
 \includegraphics[width=9.0cm]{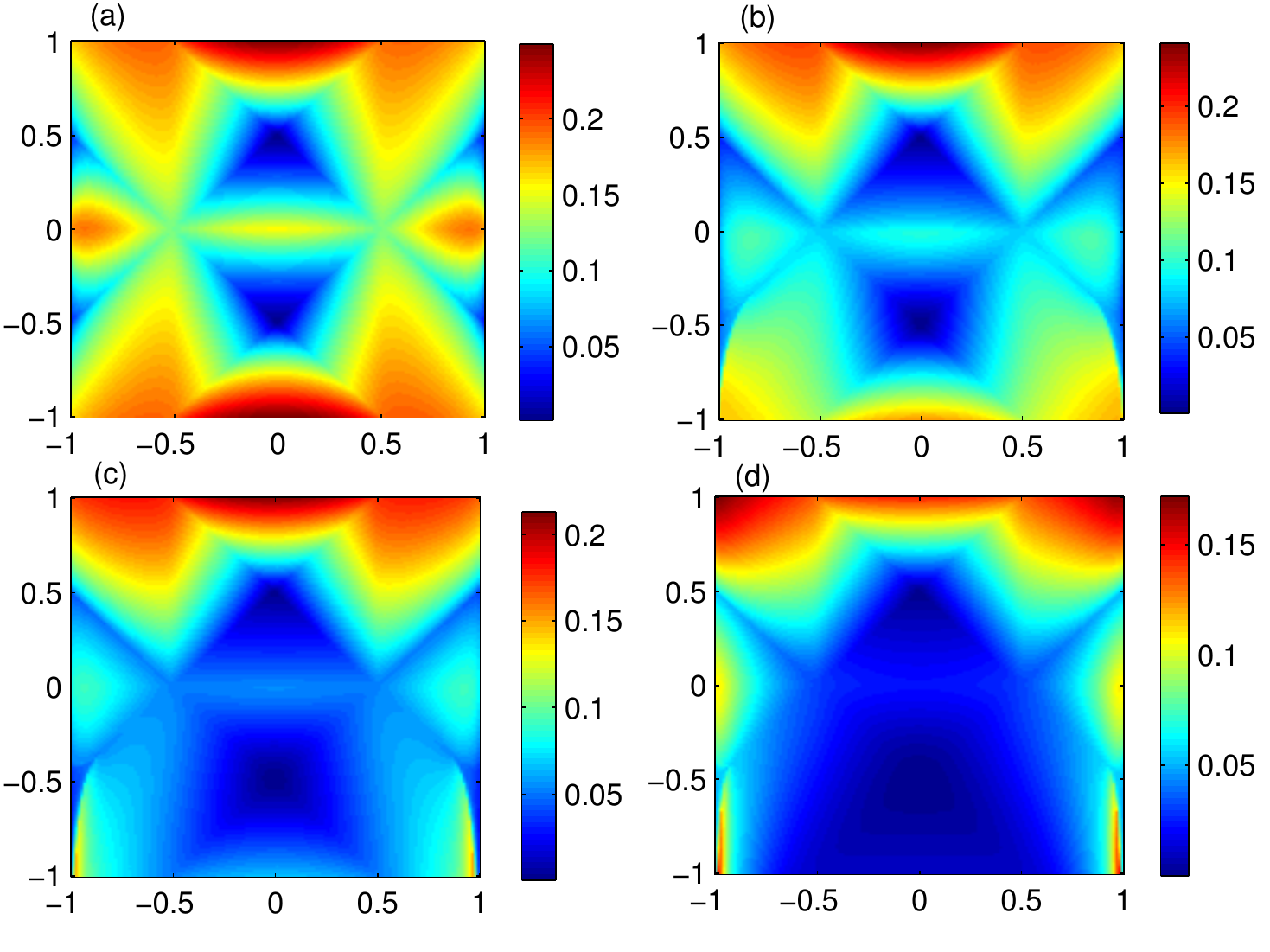}
 \caption{Conductance spectra for the choice of parameters
 $U_0=0$, $U_0=\De$, $U_0=2\De$, and $U_0=5\De$. $G h v/ W$ in units of 
 $\tb 2 e^2/h$ is plotted. The $x$-axis is $eV_{bias}/\De$ and the $y$-axis is $\mu/\De$.
 Parameters: $\De=0.003\tb$ and $E_{z0}=0.5\Delta$.}~\label{fig-U0}
\end{figure}

\begin{figure}[h]
\includegraphics[width=9.0cm]{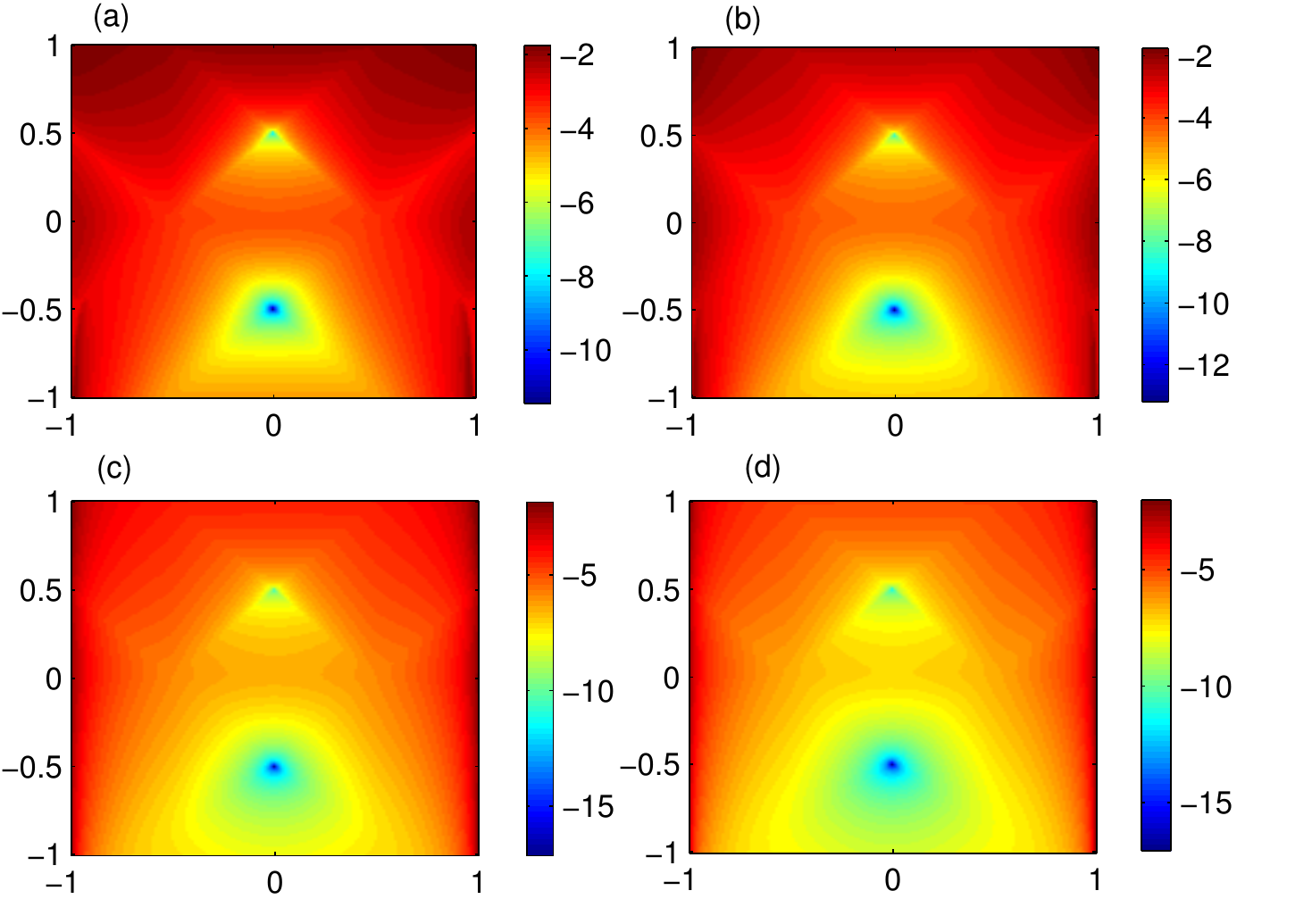}
 \caption{Conductance spectra on a logarithmic scale for the choice of parameters
 $U_0=0$, $U_0=\De$, $U_0=2\De$, and $U_0=\tb$. $\log[G h^2 v/ (W \tb 2 e^2)]$ is plotted.
 The $x$-axis is $eV_{bias}/\De$ and the $y$-axis is $\mu/\De$.
 Parameters: $\De=0.003\tb$ and $E_{z0}=0.5\De$.}~\label{fig-U0-log}
\end{figure}

The step height $U_0$ essentially captures the junction transparency. For larger magnitudes 
of $U_0$, the junction is less transparent and has a high resistance. We can see from 
Fig.~\ref{fig-U0} that for larger values of $U_0$, the features of crossover from retro- 
to specular- Andreev reflection  discussed earlier get blurred. From 
the works of Efetov et~al.~\cite{efet16,efet16-prb}, we note that when NbSe$_2$ is used as the
superconductor on top of the BLG, the parameters are $U_0$=5 meV and $\De$=1.2 meV. This 
closely corresponds to Fig.~\ref{fig-U0}~(d) and we see that the features of the 
crossover from retro- to specular- Andreev reflection begin to vanish for the
value of $U_0=5\De$. To see the features for higher values of $U_0$, we plot the conductance
on a logarithmic scale in Fig.\ref{fig-U0-log}. We see that the features discussed 
earlier vanish smoothly  over the values of $U_0=5\De, 10\De, 100\De$, and $\tb$, except for two dips at 
$(eV_{bias},\mu)=(0,\pm E_{z0})$. However, the dips correspond to orders of magnitude
smaller conductance. Thus, we find that a transparent junction is very crucial to 
observing the features of crossover from retro- to specular- Andreev reflection.

\section{Comparative analysis of the results replacing the superconductor
with normal metal}~\label{sec-scnm}
\begin{figure}
 \includegraphics[width=4.25cm]{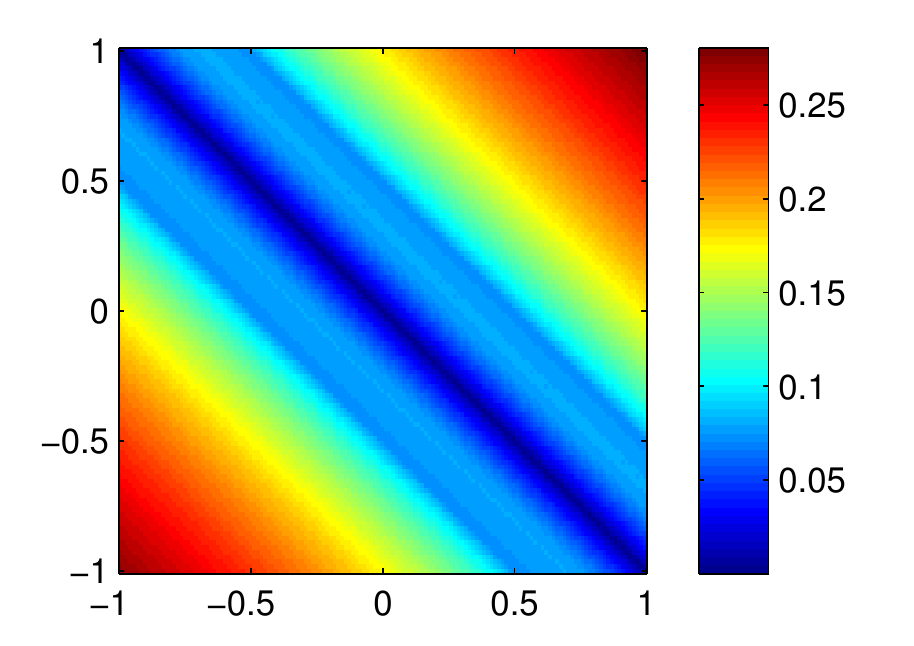}
 \includegraphics[width=4.25cm]{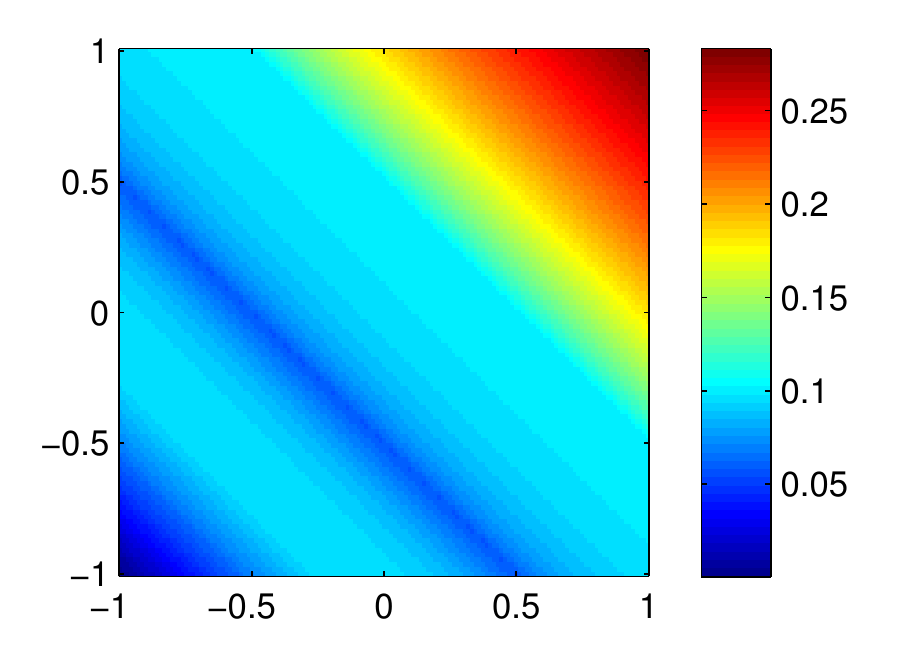}
 \caption{Conductance spectra  $G  h v/ W$ in units of $\tb 2 e^2/h$ 
 for NF junction. Left,~$U=0$, right,~$U=2\De$. The $x$-axis is 
 $eV_{bias}/\De$ and the $y$-axis is $\mu/\De$. For both, $\De=0.003\tb$ and
 $E_{z0}=0.5\De$.}~\label{fig-nf-contour}
\end{figure}

In this section, we discuss the results of the system, where superconductivity 
in the system is absent, and make comparison to the results with the system 
containing superconductivity. We denote the part of the system having a
nonzero Zeeman field by F~(ferromagnet), and N refers to the 
normal metal part which has no Zeeman field. $\De(x)=0$ for all $x$ in the NF junction. 
The calculation for the NF junction 
is presented in Appendix~A. As can be seen from the calculations, the bias $eV_{bias}$
and the chemical potential~$\mu$ enter the equations as $(eV_{bias}+\mu)$. Hence, 
the conductance depends only on the linear combination $(eV_{bias}+\mu)$ in 
the contour plot which is apparent in Fig.~\ref{fig-nf-contour}. 

\begin{figure}
 \includegraphics[width=8.5cm]{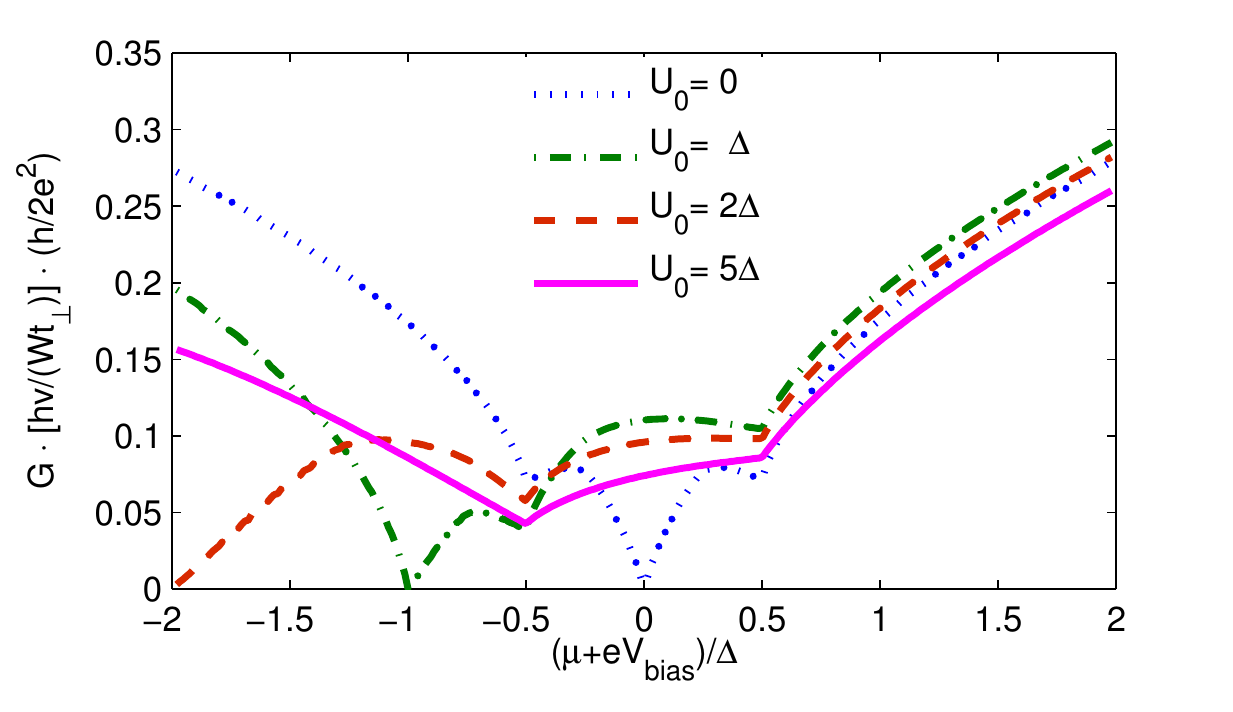}
 \caption{Conductance of NF junction, for different values of the 
 step height $U_0$. Parameters: $E_{z0}=0.5\Delta$, and
 $(eV_{bias}-\mu)=0$, where  $\De=0.003\tb$.}~\label{fig-nf-line}
\end{figure}

In Fig.~\ref{fig-nf-line}, the conductance is plotted as a function  
of $(eV_{bias}+\mu)$, for different values of step height $U_0$. 
For $U_0=0$, the conductance goes to zero at $(\mu+eV_{bias})=0$, 
since the size of the Fermi surface on the normal metal side goes to zero, 
and there are no momentum modes to carry the current. For finite values 
of $U_0$, the situation changes since at $(eV_{bias}+\mu)=0$, 
the Fermi surface has a finite size, and the current can flow from 
the F-side to the N-side. The asymmetry around 
$(\mu+eV_{bias})=0$ is because of a finite value of $U_0$. 

\begin{figure}
 \includegraphics[width=9cm]{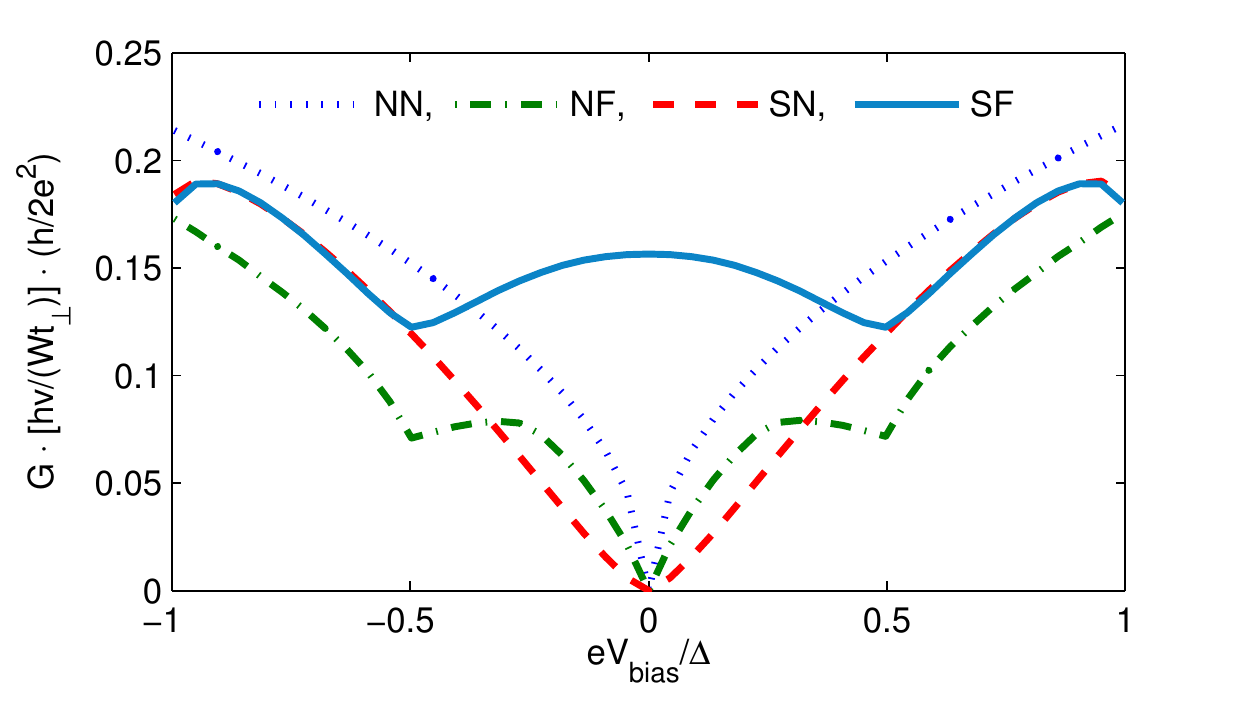}\\
 \includegraphics[width=9cm]{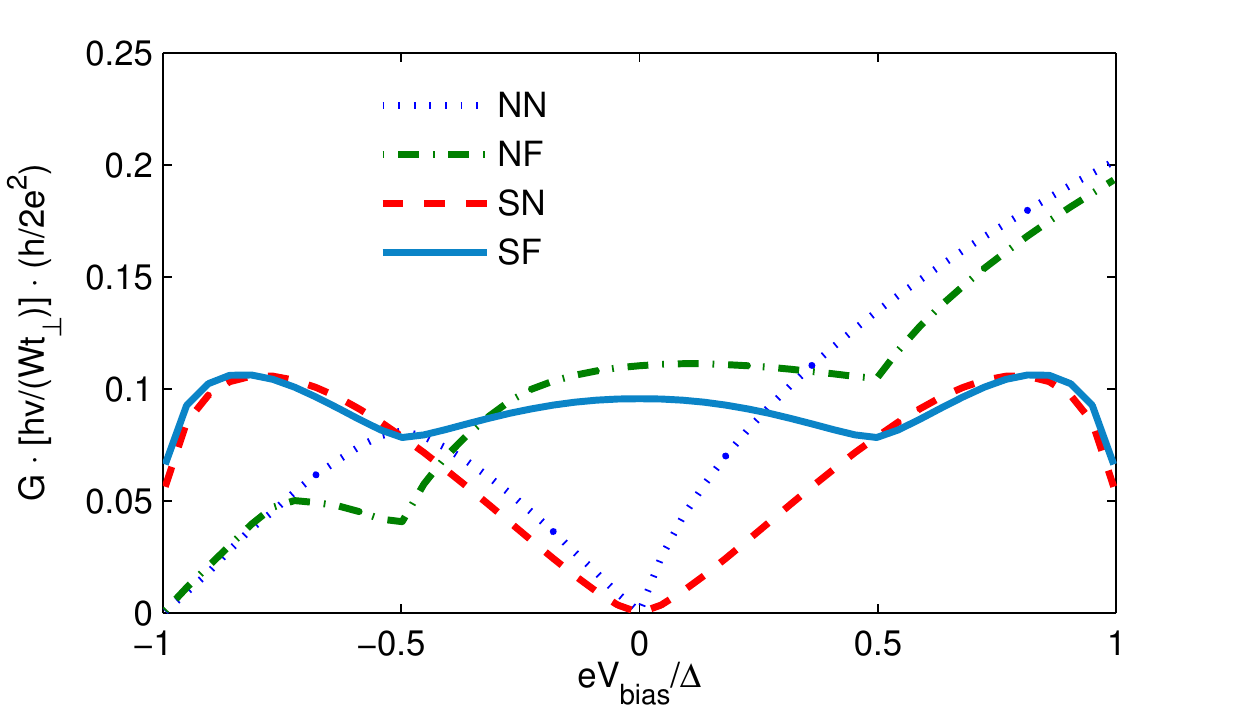}
 \caption{Conductance $G h v/ W$ in units of 
 $\tb 2 e^2/h$ is plotted for different configurations 
 of the setup: normal-normal~(NN), normal-ferromagnet~(NF),
 superconductor-normal~(SN),  and superconductor-ferromagnet~(SF).
 See text for further information. 
 Top:~$U_0=0$, bottom:~$U_0=\De$. Parameters: $E_{z0}=0.5\Delta$
 (for F), where  $\De=0.003\tb$ and $\mu=0$ (for all curves).}
 ~\label{fig-all-systems}
\end{figure}

Now, we turn to the comparison of conductances of different systems 
(NN, NF, SN, and SF) for a given choice of $U_0$ and other parameters. 
For $U_0 =0$ (Fig.~\ref{fig-all-systems},~top), 
all the curves are symmetric, while for $U_0 = \De$ the curves are 
not symmetric (except for SN and SF). For { SF}, the minima at 
$eV_{bias}=\pm E_{z0}$ and maximum at $eV_{bias}=0$ are due to 
the dispersions displaced due to Zeeman fields. This bump is where 
the specular Andreev reflection is enhanced  by the Zeeman field. 
For { NN, NF, and SN} in the case $U_0=0$, 
the conductance is zero at $eV_{bias}=0$, which is due to zero 
size of the Fermi surface of the N region. When $U_0=\De$, 
(see  Fig.~\ref{fig-all-systems},~bottom)
the size of Fermi surface is nonzero in the N region to the left 
in the { NN and NF} configurations, and there is a finite conductance
even at $eV_{bias}=0$ for the NF configuration. 

Now, we compare different curves in the bottom panel of  
Fig.~\ref{fig-all-systems}. For NN and SN configurations, 
the N-region for $x>0$ has zero sized Fermi surface at zero bias. 
Hence the conductance at  zero bias is zero (despite a nonzero sized Fermi
surface in the region $x<0$ for NN). Now, when we turn to the case of NF, 
the Fermi surfaces on both sides of the junction at $eV_{bias}=0$ have 
nonzero size. Hence, the conductance is finite around $eV_{bias}=0$. The 
conductance for NF approaches zero as $eV_{bias}\to \De$ since the size of 
Fermi surface approaches  zero on the N-side of the junction as 
we have chosen $U_0=\De$. For the case of SF, the conductance is nonzero 
in the entire range shown since the size of the Fermi surface on F-side 
is always nonzero due to a finite value of the Zeeman field ($E_{z0}=0.5\De$), 
and on the {\bf S}-side there is superconducting gap which favors Andreev
reflection. Finally, the conductances in the lower panel are  smaller
than those in the upper panel since the step height $U_0$ is zero in the
upper panel and is $\De$ in the lower panel, reducing the transparency 
of the junctions studied in the lower panel. 

\section{Experimental relevance}~\label{sec-expt}
To implement our scheme experimentally, it is important to apply a 
Zeeman field in the NM part of the junction. An in-plane magnetic field which is 
less than the critical field to kill the superconductivity of the SC 
part in the system will achieve this. Another way to implement a Zeeman field
is to bring a ferromagnetic insulator in proximity to the NM-side of the 
junction. It has been shown that ferromagnetism can be induced in 
graphene by such proximity coupling with several materials such  as 
EuO, YIG and EuS~~\cite{haugen08,wang15,wei16}. 

A typical sample will have a
disorder which manifests as Fermi energy broadening $\delta \ep_F$. This means that
the BLG sample must be of a sufficiently high quality so that the  Fermi energy
broadening $\delta \ep_F$ is small ($\delta \ep_F \ll \De$). Furthermore, observing 
the features of crossover for a fixed bias $eV_{bias}\ll\De$ as 
$\mu$ is varied is important as the quasiparticle contribution to transport
is the least in this regime. In addition, a finite temperature will result in
thermal broadening and hence, performing the experiment at a low temperature is 
necessary to observe the features discussed here. The temperature has to be low compared to 
both the superconducting gap~($\sim 14~K$ in NbSe$_2$~~\cite{efet16}) and Zeeman energy~($\sim 10~K$). Experimentally, 
reaching temperatures of about $100~mK$ is possible and hence temperature does not pose a 
hindrance to implementing our scheme in realistic systems. 

In a realistic system, the work function mismatch
between the NM and SC regions can result in the formation of a NM region having a 
length-scale $a$ at the interface as discussed in Ref.~\cite{tanake17}. 
Also, from the value of the work functions of NbSe$_2$ and 
BLG, the step height $U_0$ is chosen to be $1eV>\tb$ in Ref.~\cite{tanake17} in 
contrast to the limit $U_0\ll\tb$ in Ref.~\cite{efet16,efet16-prb} where the value of 
$U_0$ is chosen to match the experimental results. Our calculations combined 
with the choice of $U_0$ in Ref.~\cite{efet16,efet16-prb} point to a small value of 
$a$ ($a\ll 100 nm$) in contrast to the assertion made in 
Ref.~\cite{tanake17}. This means that the effects of a p-n junction formed
at the NM-SC interface may be negligible. In Appendix~B,
we study the effect of having a finite $a$ and show that it can be negligible. 

\section{Summary and conclusion}~\label{sec-summary}
We have studied Andreev reflection at a junction of bilayer graphene 
and a superconductor.  Since our main objective has been to observe the enhanced signatures of
specular Andreev reflection, we introduce a Zeeman field and study the features 
on a contour plot of conductance versus chemical potential and bias voltage when
these two energy scales are  less than the superconducting gap. We find that a finite
Zeeman field produces a diamond shaped region at the center where the Andreev 
reflection is purely specular. Furthermore, the lines bordering the diamond shaped region 
and two patches around the low bias region at the corners of the diamond show 
a low conductance, where the crossover from specular- to retro- type Andreev
reflection occurs. Importantly, we find that for a barrier step-height that is of the same order of 
magnitude as the superconducting gap, the features of the crossover from 
retro- to specular- Andreev reflection are observable and for a barrier step-height 
much larger than the superconducting gap, the features vanish except for small 
regions of low conductance at $(eV_{bias},\mu)=(0,\pm E_{z0})$. We have also analyzed the 
relative contributions from normal state conductance, where the superconductivity is switched 
off. Furthermore, we have discussed how our calculations can be tested in an experimental system. 

\acknowledgments AD thanks Nanomission, Department of Science and Technology (DST) 
for the financial support under grants - DSTO1470 and DSTO1597. 
AS thanks DST Nanomission (DSTO1597) for funding. SM thanks the Indo-Israeli UGC-ISF project for funding.

\section*{Appendix A}
In this section, we give details of the calculation for the system 
comprising of a Zeeman field induced ferromagnetic region in contact with 
the normal metal region. This is simply the limit of the NM-SC junction described
by Eq.~\eqref{ham} where $\De(x)=0$ for all $x$. The wavefunction for an 
electron incident on the junction  from $x>0$ onto $x<0$, with energy $E$ has the form 
$\phi_s(x)e^{i k_y y}$, where 
\bea 
\phi_s(x) &=& e^{-ik^e_xx}~ \vec u_{N,s}(\ep,-k^e_x) + r_N~ e^{ik^e_xx}
~\vec u_{N,s}(\ep,k^e_x) \nn \\ && + ~\tilde r_N ~e^{-\kappa x}~\vec u_{N,s}(\ep,i\kappa), 
{\rm ~~for~~} x>0 \nn \\ 
&=& t_N e^{-i\tilde k^e_xx}~ \vec u_{N,s}(\tilde \ep,-\tilde k^e_x) 
+  ~\tilde t_N e^{\tilde \kappa x}~\vec u_{N,s}(\tilde \ep,-i\tilde \kappa) \nn \\
&& {\rm ~~for ~~} x<0.
\eea 
Here, $\ep=E+\mu+sE_{z0}$, $\tilde\ep=E+\mu+U_0$, 
$k_y=\sqrt{\ep^2+\tb|\ep|}\sin{\th}/(\hbar v)$ ($\th$ is the angle of incidence so that
the normal incidence corresponds to $\th=0$), 
\bea
k^e_x&=&sign(\ep) \sqrt{\ep^2+\tb|\ep|-(\hbar v k_y)^2}/(\hbar v), \nn \\
\kappa &=& \sqrt{(\hbar v k_y)^2 + \tb|\ep|-\ep^2}/(\hbar v), \nn \\
\tilde k^e_x &=& sign(\tilde  \ep)\sqrt{\tilde\ep^2+\tb|\tilde\ep|-(\hbar v k_y)^2}/(\hbar v),\nn \\
{\rm and~~} \tilde \kappa &=& \sqrt{(\hbar v k_y)^2 + \tb|\tilde\ep|-\tilde\ep^2}/(\hbar v).\nn \\
\eea
Now, using the boundary condition, which is continuity of the wavefunction at $x=0$, 
one can determine the scattering amplitudes $r_N$, $\tilde r_N$, $t_N$, and $\tilde t_N$. 
With this, the wavefunction is determined and using a formula similar to Eq.~\eqref{eq-conductance}, 
the conductance can be calculated. 

\section*{Appendix B}
In this part, we study the effect of having a finite region of length 
$a$ on the NM part of the junction where $U(x)\neq 0$. The Hamiltonian 
has the same form as in Eq.~\eqref{ham}, except for two changes:
$U(x)=U_0\eta(a-x)$ and $E_z(x)=E_{z0}\eta(x-a)$, where $\eta(x)$
is a Heavyside step function.
The wavefunction for an electron at energy $E$
(in the range: $|E|<\De\ll\tb$) and spin~$s$ 
($s=\pm1$ is the eigenvalue of the operator $s_z$), incident from 
the NM side onto the SC has the form $\psi_s(x)e^{ik_y y}$, such that  

\bea  
\psi_s(x) &=& \Big(e^{-ik^e_xx}~ \vec u_{N,s}(\ep,-k^e_x) + r_N~ e^{ik^e_xx}
~\vec u_{N,s}(\ep,k^e_x)\Big)  \begin{bmatrix}
                          1 \\0
                         \end{bmatrix} \nn \\
& & + ~r_A ~e^{-i k^h_x x}~\vec v_{N,s}(\ep_h,-k^h_x) \begin{bmatrix}
                                             0 \\ 1
                                            \end{bmatrix} \nn \\
& &    + ~\tilde r_N ~e^{-\kappa x}~\vec u_{N,s}(\ep,i\kappa) \begin{bmatrix}
                                             1 \\ 0
                                            \end{bmatrix} \nn \\
& & + ~\tilde r_A ~e^{-\ka^h x}~\vec v_{N,s}(\ep_h,i\ka^h) \begin{bmatrix}
                                             0 \\ 1
                                            \end{bmatrix},~~
                                            {\rm for}~~x>a, \nn \\ 
&=& \Big( s_{e-}~e^{-ik^{e'}_x x} \vec u_{N',s}(\ep',-k^{e'}_x) \nn \\ 
 && + s_{e+}~e^{ik^{e'}_x x} \vec u_{N',s}(\ep',k^{e'}_x)\Big)  \begin{bmatrix}
                          1 \\0
                         \end{bmatrix} 
\nn \\ 
&& + \Big( s_{h-}~e^{-ik^{h'}_x x} \vec v_{N',s}(\ep',-k^{h'}_x) \nn \\ 
 && + s_{h+}~e^{ik^{h'}_x x} \vec v_{N',s}(\ep',k^{h'}_x)\Big)  \begin{bmatrix}
                          0 \\ 1
                         \end{bmatrix} 
\nn \\ 
&&+ \Big( \tilde s_{e-}~e^{-\ka^{e'}_x x} \vec u_{N',s}(\ep',i\ka^{e'}_x) \nn \\ 
 && + \tilde s_{e+}~e^{\ka^{e'}_x x} \vec u_{N',s}(\ep',-i\ka^{e'}_x)\Big)  \begin{bmatrix}
                          1 \\0
                         \end{bmatrix} 
\nn \\ 
&& + \Big( \tilde s_{h-}~e^{-\ka^{h'}_x x} \vec v_{N',s}(\ep',i\ka^{h'}_x) \nn \\ 
 && + \tilde s_{h+}~e^{\ka^{h'}_x x} \vec v_{N',s}(\ep',-i\ka^{h'}_x)\Big)  \begin{bmatrix}
                          0 \\ 1
                         \end{bmatrix} {\rm,~for}~~0<x<a,
\nn \\ 
&=& \sum_{j=1}^4 w_{j,s}~  e^{i k^S_j x}~ \vec u_{S}
(k^{S}_j),~~ {\rm for}~~x<0, \eea
where $\vec u_{N,s}(\tilde\ep,k_x)$ and $\vec v_{N,s}(\tilde\ep,k_x)$ are 
the electron- and hole- sector eigenspinors of the Hamiltonian on
the NM side [given by Eq.~\eqref{eigsp}] with $x$-component of momentum $k_x$,
and $\vec u_S(k^S_j)$ is the eigenspinor
on the SC side with $x$-component of  momentum~$k^S_j$. Furthermore,  
the $x$-component of electron and hole momenta on the NM side are given by:
\bea
\hbar v k^e_x &=& sign(\ep) \sqrt{\ep^2+\tb|\ep|-(\hbar v k_y)^2}, \nn \\
\hbar v k^h_x &=& sign(\ep_h) \sqrt{\ep^2_h+\tb|\ep_h| -(\hbar v k_y)^2}, \nn \\
\hbar v \ka &=& \sqrt{(\hbar v k_y)^2 + \tb|\ep|-\ep^2}, \nn \\
\hbar v \ka^h &=& \sqrt{(\hbar v k_y)^2 + \tb|\ep_h|-\ep_h^2} ,\nn \\
\hbar v k^{e'}_x &=& sign(\ep') \sqrt{\ep'^2+\tb|\ep'|-(\hbar v k_y)^2}, \nn \\
\hbar v k^{h'}_x &=& sign(\ep'_h) \sqrt{\ep'^2_h+\tb|\ep'_h| -(\hbar v k_y)^2}, \nn \\
\hbar v \ka_x^{e'} &=& \sqrt{(\hbar v k_y)^2 + \tb|\ep'|-\ep'^2}, \nn \\
\hbar v \ka_x^{h'} &=& \sqrt{(\hbar v k_y)^2 + \tb|\ep'_h|-{\ep'}_h^2} ,
\eea
where $\ep = (E+\mu+sE_{z0})$, $\ep_h=(\mu-sE_{z0}-E)$, $\ep' = (E+\mu+U_0)$ and 
$\ep'_h=(\mu+U_0-E)$. The continuity of $\psi_s(x)$ at $x=0$ and $x=a$ in total give 
16 equations for 16 scattering amplitudes to be solved. Then, the conductance is  
calculated using Eq.~\eqref{eq-conductance}. 

\begin{figure}
  \includegraphics[width=8.4cm]{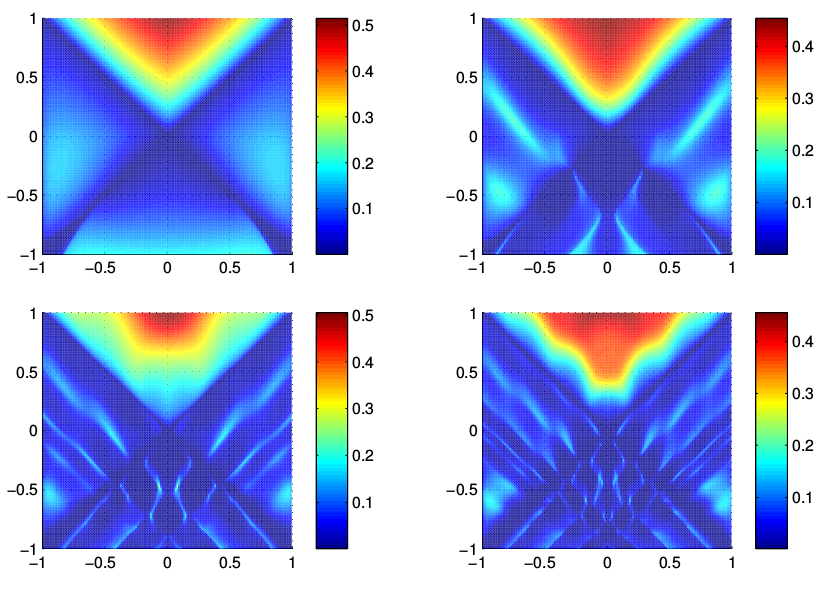}
 \caption{Conductance spectra for the choice of parameters
 $a=10,50,100,150$ (in units of $\hbar v/\tb$) for top-left, top-right, 
 bottom-left, bottom-right respectively.  
 $x$-axis is $eV_{bias}/\De$ and $y$-axis is $\mu/\De$.
 Parameters: $\De=0.003\tb$, $U_0=\De$ and
 $E_{z0}=0$.}~\label{fig-ez0-lam1}
\end{figure}

First, the conductance is calculated for $E_{z0}=0$, for various values of $a$ 
and a fixed value of $U_0=\De$ in Fig.~\ref{fig-ez0-lam1}. It can be seen that 
for higher values of $a$, Fabry-P\'erot type oscillations~\cite{soori12} 
are observed in the conductance spectra. Comparing this with 
the experimental results in Ref.~\cite{efet16}, the absence of conductance 
oscillations there suggests that in a realistic system, $a$ is 
small ($a\ll50\hbar v/\tb$). 

\begin{figure}
  \includegraphics[width=8.4cm]{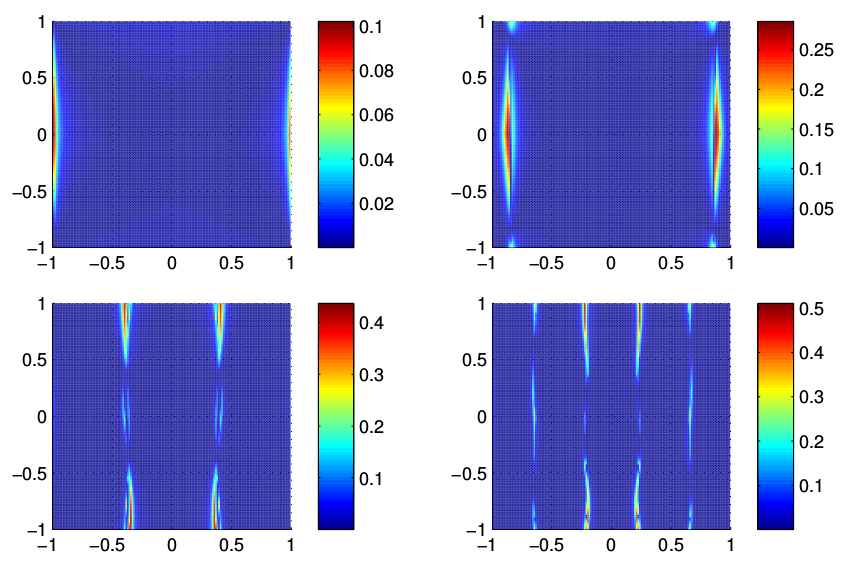}
 \caption{Conductance spectra for the choice of parameters
 $a=10,100,500,$ and $1000$ (in units of $\hbar v/\tb$) for top-left, top-right, 
 bottom-left, and bottom-right panels respectively.  
The $x$-axis is $eV_{bias}/\De$ and the $y$-axis is $\mu/\De$.
 Parameters: $\De=0.003\tb$, $U_0=5\tb$ and
 $E_{z0}=0$.}~\label{fig-ez0-lam2}
\end{figure}

Next, we study the case of $U_0=5\tb$ (discussed in Ref.~\cite{tanake17})
keeping $E_{z0}=0$ in Fig.~\ref{fig-ez0-lam2} for different values of $a$.
We see that for larger values of $a$ ($a>100\hbar v/\tb$), 
there are Fabry-P\'erot type oscillations in conductance. Comparing these 
with the experimental results in Ref.~\cite{efet16}, we see that $a$ must 
be small ($a\ll 100\hbar v/\tb$). While the precise values of $U_0$ and $a$
are unknown in a realistic system, our results suggest that 
$U_0\sim \De$ and $a\lesssim10\hbar v /\tb$. Furthermore, this limit of $U_0$
and $a$ is important to observe the features of the crossover from retro 
to specular Andreev reflection in a system with finite $E_{z0}$. 

\begin{figure}
  \includegraphics[width=8.40cm]{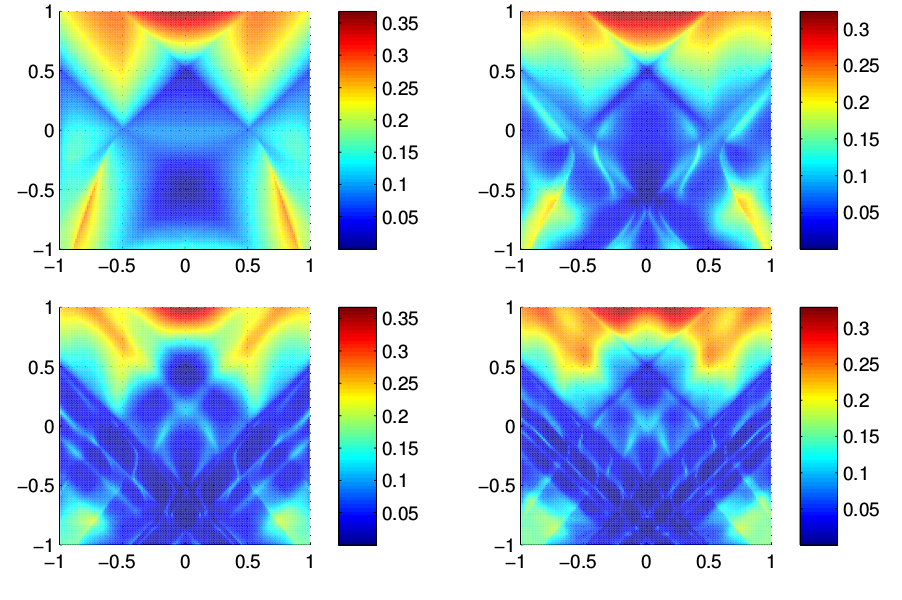}
 \caption{Conductance spectra for the choice of parameters
 $a=10,50,100,$ and $150$ (in units of $\hbar v/\tb$) for top-left, top-right, 
 bottom-left, and bottom-right panels respectively.
  The $x$-axis is $eV_{bias}/\De$ and the $y$-axis is $\mu/\De$.
 Parameters: $\De=0.003\tb$, $U_0=\De$ and
 $E_{z0}=0.5\De$.}~\label{fig-ez-lam1}
\end{figure}

Now, we turn to the case of $E_{z0}=0.5\De$. In Fig.~\ref{fig-ez-lam1}, 
we see how the conductance spectrum changes as $a$ is changed keeping 
$U_0=\De$ fixed. The features of crossover still remain, but there 
are oscillations in the conductance spectrum due to Fabry-P\'erot type 
interference, which occur due to 
modes in the region $0<x<a$. The two dark regions of low conductance
around the  points $B$ and $D$, and the dark lines $PD$ and $DS$ remain.
Furthermore, the dark lines $BA$ and $BC$ remain, while the dark lines along 
$AQ$ and $CS$ vanish. It is not possible to distinguish the Fabry-P\'erot
oscillations in the conductance spectrum from the crossover from specular 
to retro Andreev reflection, but with a knowledge of $E_{z0}$ and $\De$
the points: $P,Q,R,S,A,B,C,$ and $D$ in the conductance spectrum can be identified,
thereby finding the crossover lines.

\end{document}